\documentclass[aps,pre,twocolumn,superscriptaddress,floats,floatfix,showpacs]{revtex4-1}
\usepackage{amssymb,amsmath}
\usepackage{float}
\usepackage{mathrsfs}
\usepackage{graphics}
\usepackage{epstopdf}
\usepackage{color}
\usepackage{subfigure}
\usepackage{epsfig}
\usepackage{rotating}
\usepackage{anyfontsize}
\usepackage{verbatim}
\usepackage{standalone}
\usepackage{hyperref}
\usepackage{array}
\usepackage{xr}

\begin{document}

\title{Arrhythmogenicity of cardiac fibrosis: fractal measures  
and Betti numbers}
\author{Mahesh Kumar Mulimani}
\email{maheshk@iisc.ac.in ;}
\affiliation{Centre for Condensed Matter Theory, Department of Physics, Indian Institute of Science, Bangalore 560012, India.}
\author{Brodie A. J. Lawson}
\email{brodie.lawson86@gmail.com ;}
\affiliation{Centre for Data Science, Queensland University of Technology, Brisbane, Australia.}
\affiliation{ARC Centre of Excellence for Mathematical and Statistical Frontiers, Queensland University of Technology, Brisbane, Australia.}
\author{Rahul Pandit}
\email{rahul@iisc.ac.in}
\altaffiliation[\\]{also at Jawaharlal Nehru Centre For
Advanced Scientific Research, Jakkur, Bangalore, India}
\affiliation{Centre for Condensed Matter Theory, Department of Physics, Indian Institute of Science, Bangalore 560012, India.}


\begin{abstract}

Infarction- or ischaemia-induced cardiac fibrosis can be arrythmogenic.  We use
mathematcal models for  diffuse fibrosis ($\mathcal{DF}$), interstitial
fibrosis ($\mathcal{IF}$), patchy fibrosis ($\mathcal{PF}$), and
compact fibrosis ($\mathcal{CF}$) to study patterns of
fibrotic cardiac tissue that have been generated by new mathematical
algorithms.  We show that the fractal dimension $\mathbb{D}$, the
lacunarity $\mathcal{L}$, and the Betti numbers
$\beta_0$ and $\beta_1$ of such patterns are \textit{fibrotic-tissue markers}
that can be used to characterise the arrhythmogenicity of different
types of cardiac fibrosis. We hypothesize, and then demonstrate by
extensive \textit{in silico} studies of detailed mathematical models
for cardiac tissue, that the arrhytmogenicity of fibrotic tissue is
high when $\beta_0$ is large and the lacunarity parameter $b$ is small.

\end{abstract}

\maketitle

Sudden cardiac death (SCD) continues to be a leading cause of death in the
industrialised world (see, e.g.,
Refs.~\cite{kawara2001activation,biernacka2011aging,nguyen2014cardiac,hinderer2019cardiac}
and ~\cite{SCD_stats}).
Even young athletes~\cite{stormholt2021symptoms} may be victims of SCD; and a recent
study has suggested that there is a correlation between out-of-hospital cardiac
arrest and COVID-19 ~\cite{baldi2020out,kuck2020arrhythmias}. Ventricular arrhythmias, such as ventricular tachycardia
(VT) and ventricular fibrillation (VF), are often the root cause of
SCDs~\cite{mehra2007global,rubart2005mechanisms}. Myocardial infarction and ischaemia lead to \textit{cardiac-tissue
fibrosis}, which is one of the important contributors to arrhythmogenesis, and,
therefore, to SCDs.  Several experimental studies, such as those in
Refs.~\cite{kawara2001activation,hocini2002electrical,balaban2018fibrosis},
have demonstrated the arrythmogenicity of cardiac fibrosis, which induces
reentry by delaying local conduction. Fibrosis has been observed to alter
the dynamics of the electrical waves passing through fibrotic regions; this leads
to the formation of re-entrant waves that can precipitate cardiac
arrythmias~\cite{majumder2012nonequilibrium,morgan2016slow,jousset2016myofibroblasts,clayton2018dispersion}.

In the heart, fibrotic tissue is made up of cardiac fibroblast cells or
collagen fibers; and it has been classified
visually~\cite{nguyen2014cardiac,hansen2017fibrosis} into four different types
with (a) diffuse fibrosis ($\mathcal{DF}$), (b) interstitial fibrosis
($\mathcal{IF}$), (c) patchy fibrosis ($\mathcal{PF}$), and (d) compact
fibrosis ($\mathcal{CF}$).  However, \textit{in-vivo}, \textit{ex-vivo}, or
\textit{in-vitro} studies have not been used hitherto for a quantitative
statistical characterization of these types of fibrosis, perhaps because
large-enough data sets of images are not available. We show, via detailed
analysis, that recently developed mathematical models~\cite{jakes2019perlin} for
fibrotic tissue, which use Perlin noise, and idealised models, which we define
below, can be used to distinguish quantitatively between $\mathcal{DF}$,
$\mathcal{IF}$, $\mathcal{CF}$, and $\mathcal{PF}$ by obtaining the fractal
dimension $\mathbb{D}$, the lacunarity $\mathcal{L}(\epsilon)$, and Betti
numbers $\beta_0$ and $\beta_1$ (see, e.g.,
Refs.~\cite{delacalleja2020fractal,gould2011multifractal}) of patterns of
fibrotic tissue. For fibrosis patterns, which we obtain from Perlin noise, we employ the 
notations $\mathcal{DFP}$, $\mathcal{IFP}$, $\mathcal{CFP}$, and $\mathcal{PFP}$ 
for diffuse, interstitial, patchy, and compact fibrosis, respectively;
their counterparts for the idealised model are $\mathcal{DFI}$,
$\mathcal{IFI}$, $\mathcal{CFI}$, and $\mathcal{PFI}$.
We show how to compute such properties by the digitisation of
images of fibrotic tissue. These properties serve as \textit{fibrotic-tissue
markers}; and they can be used to characterise the arrhythmogenicity of
different types of cardiac fibrosis. We hypothesize, and then demonstrate by
extensive \textit{in silico} studies of detailed mathematical models for
cardiac tissue, that the arrhythmogenicity of fibrotic tissue is high when
$\beta_0$ is large and the lacunarity parameter $b$ is small. Our study has
implications for clinical cardiology, because, even at a qualitative level, we
find that (a) $\mathcal{DF}$ is most arrythogenic and (b) $\mathcal{CF}$ is
least arrythmogenic.

For the dynamics of cardiac myocytes we use the biologically realistic
human-ventricular-cell model~\cite{ten2006alternans}, due to ten Tusscher and
Panfilov (henceforth, the TP06 model), in which the spatiotemporal evolution of the
transmembrane potential $V_m$ is governed by the following reaction-diffusion
partial differential equation (PDE):
\begin{equation}
\frac{\partial{{V_m}}}{\partial{t}}+\frac{I_{{ion}}}{C_m}=\nabla.(D\nabla {V_m});
	\label{eq:VmPDE}
\end{equation}
here, $I_{ion}$ is the sum of all the ionic currents (Eq.~\ref{eq:Iionsum}),
$C_m$ is the membrane capacitance, and, in the case of tissue with healthy
myocytes, we use a scalar diffusion constant $D=0.00154 \ cm^2/ms$; in the
region of the tissue with fibrosis and with collageneous fibers we use $D=0$.
\begin{eqnarray}
	I_{ion} &=& I_{Na} + I_{CaL}+I_{to}+I_{Ks}+I_{Kr}+I_{K1}+I_{NaCa} \nonumber \\
	&+& I_{NaK}+ I_{pCa}+I_{pK}+I_{bNa}+I_{bCa}.
\label{eq:Iionsum} 
\end{eqnarray}

For the details of the currents we refer the reader to the
Ref.~\cite{ten2006alternans}; and we use the standard TP06-model
parameters~\cite{ten2006alternans} for ion-channel conductances.  We obtain the
ordinary differential equation (ODE) for a single cardiomyocyte by setting
$D=0$ in Eq.~\ref{eq:VmPDE}. Our model for cardiac tissue has three types of
regions:
\begin{itemize}
\item those in which we have normal,
TP06-model mycoytes that evolve according to Eq.~\ref{eq:VmPDE} with  $D =
0.00154  \ cm^2/ms$; 
\item in the vicinities of fibrotic areas, there are
regions in which remodelled cardiomyocytes evolve according to
Eq.~\ref{eq:VmPDE}, with  $D = 0.00154  \ cm^2/ms$, but with modified
conductances [we change the maximal ion-channel conductances $G_{Na}, \
G_{CaL}, \ G_{Kr} \ \text{and} \ G_{Ks}$, in the TP06 model, to $0.38*G_{Na}, \
0.31*G_{CaL}, \ 0.3*G_{Kr} \ \text{and} \ 0.2*G_{Ks}$, respectively, (see,
e.g., Ref.~\cite{zlochiver2008electrotonic,mcdowell2011susceptibility,nguyen2014cardiac})];
\item fibrotic-tissue regions in which we set $D = 0$.
\end{itemize}

\begin{figure}
\center
\includegraphics[scale=0.30]{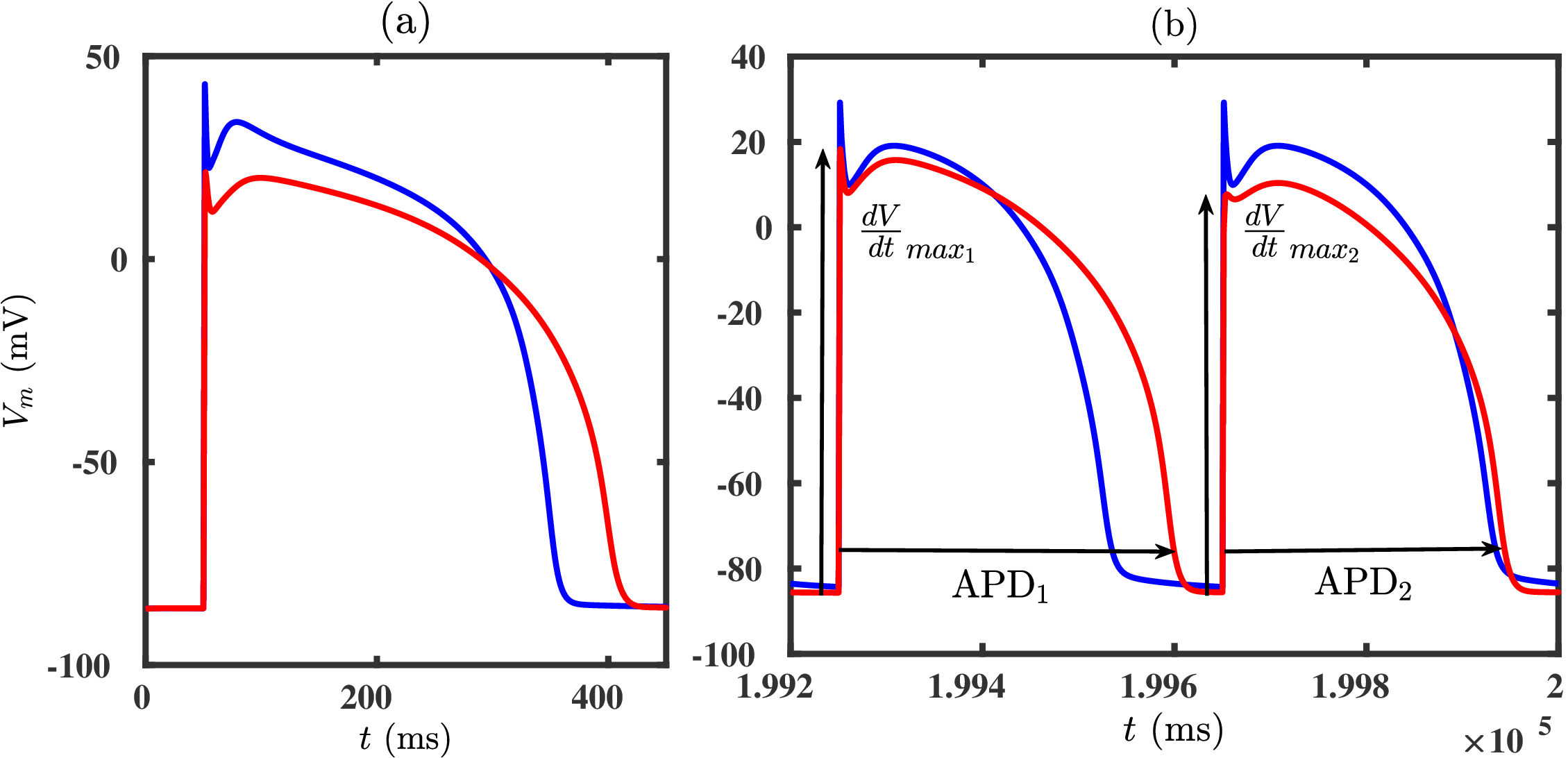}
\caption{(Color online) Plots from our simulations of APs of normal (NM) [blue] and remodelled (RM) [red] 
myocytes: (A) Both the NM and the RM are given a
single stimulus pulse; and (B) the $1000^{th}$ and $999^{th}$ APs of NM and
RM that are paced with a pacing frequency of PCL=$400$ ms ($2.5$ Hz); here,
we observe alternans for the RM but not for the NM.}
\label{fig:fig1}		
\end{figure}

First we show the effect of remodeling on a single myocyte cell. In
Fig.~\ref{fig:fig1} we contrast the action potential (AP) of a normal-myocyte
(NM) and a remodelled-myocyte (RM): The upstroke-velocity $\frac{dV}{dt}_{max}$
of the AP decreases and the action-potential duration (APD) increases if we
replace a NM by RM.  Furthermore, when we pace these myocytes, with a
pacing frequency of $2.5$ Hz, we observe alternans only in RM. We expect,
therefore, that in those parts of the tissue that have RMs the conduction
velocity of the wave decreases and its wavelength $\lambda$ increases. (For
the importance of such remodeling, see Fig. S5 in the Supplemental
Material~\cite{Supp_Mat}.)

\begin{figure}
 \center
	\includegraphics[scale=0.275]{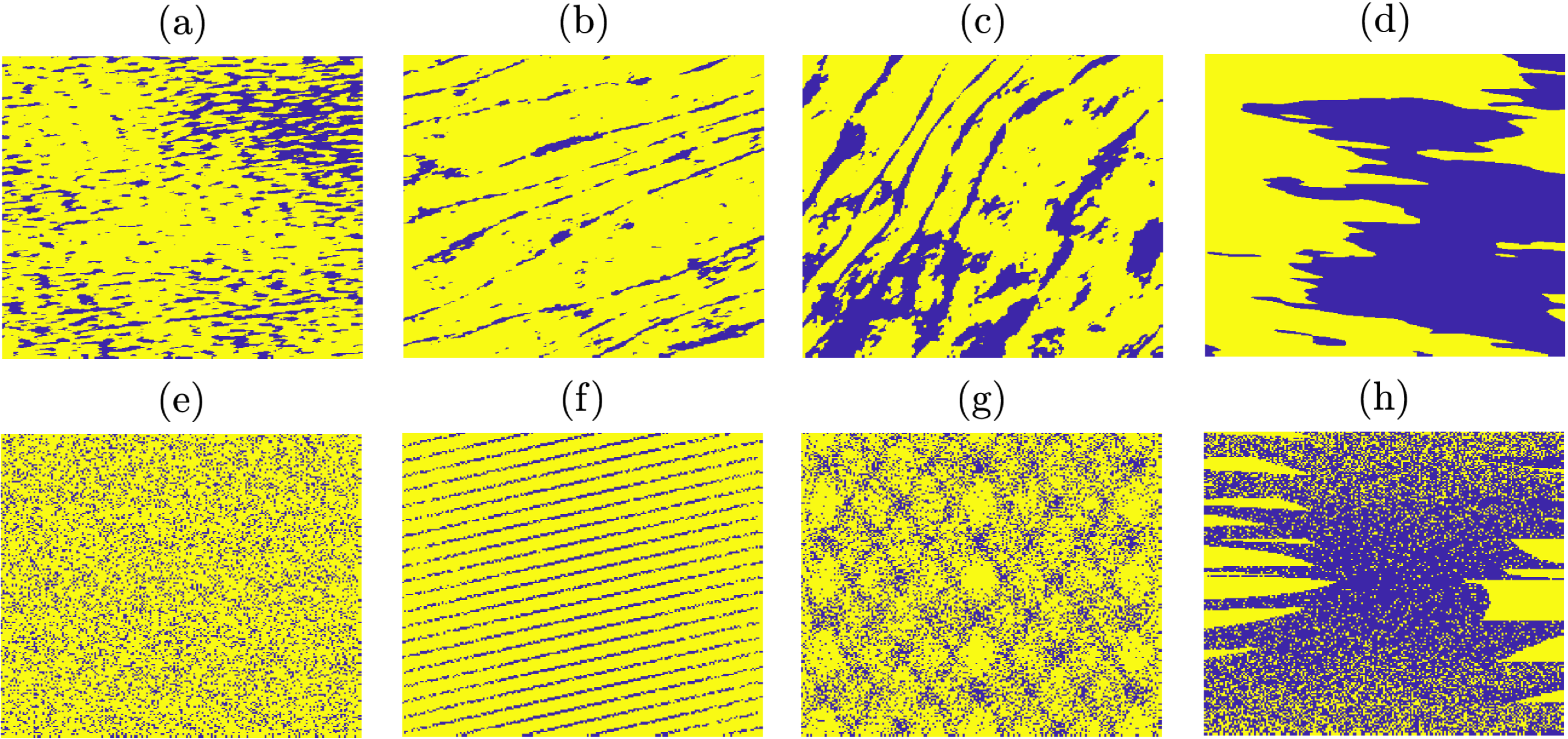}
\caption{(Color online) Illustrative plots of fibrotic textures of types (a) $\mathcal{DFP}$, 
(b) $\mathcal{IFP}$, (c) $\mathcal{PFP}$, and (d) $\mathcal{CFP}$, from the Perlin-noise
model (A) (see text and Ref.~~\cite{jakes2019perlin}); regions with myocytes and with fibrotic tissue
are indicated in yellow and blue, respectively. (e) $\mathcal{DFI}$, 
(f) $\mathcal{IFI}$, (g) $\mathcal{PFI}$, and (g) $\mathcal{CFI}$ in the second row
are the counterparts of (a)-(d) for the idealised model (B) (see text).}
\label{fig:fig2}	
\end{figure}

We use the following two classes of mathematical models for the organization of
fibrotic tissue: 
\begin{itemize}
\item (A) A recently developed model~\cite{jakes2019perlin} 
yields fibrotic textures of types $\mathcal{DF}$, $\mathcal{IF}$, 
$\mathcal{PF}$, and $\mathcal{CF}$, for which we give illustrative plots in 
Figs.~\ref{fig:fig2} (a), (b), (c), and (d), respectively, with normal (yellow) and 
fibrotic (blue) regions. This model creates synthetic textures, for different 
types of fibrotic tissue, by using Perlin noise and approximate Bayesian 
computation~\cite{jakes2019perlin}; these textures match well with 
those observed in experiments. 
\item (B) Idealised models, which we define below, in a square region
($R \times R$ grid of myocytes); these models include parameters like $p_f$, 
the percentage of fibrotic sites, and $\theta \ \in \ [0,\pi]$, the angle that 
fibrotic strands make with the horizontal axis; these parameters can be tuned easily.  
\begin{itemize}
\item (i) $\mathcal{DFI}$: we replace, randomly, a percentage $p_f$ of myocytes by 
fibrotic, nonconducting ($D=0$) sites (Fig.~\ref{fig:fig2} (e));
\item (ii) $\mathcal{IFI}$: we introduce long, thin strands of non-conducting fibers
($D=0$), with orientation $\theta$; fiber thickness: $2-3$ grid points; 
fiber lengths go from a minimum of $2$ to at most $50-60$ grid points 
(Fig.~\ref{fig:fig2} (f)).
\item (iii) $\mathcal{PFI}$: we use strands, as in $\mathcal{IF}$, but with 
two different angles for interstitial fibers, say $\theta_1= 60 ^\circ$ 
and $\theta_2=90^\circ +60^\circ $, with thicknesses and lengths of $2-3$
grid points; at the intersection of fibers, we add small 
patches of diffuse fibrosis with $p_f = 5-10 \%$ (Fig.~\ref{fig:fig2} (g)).
\item (iv) $\mathcal{CFI}$: we generate compact regions with random shapes 
(Fig.~\ref{fig:fig2} (h)) by using the pinta software (Ref.~\cite{PINTA}), a 
program for drawing.
\end{itemize}
\end{itemize}

\begin{figure}
 \center
\includegraphics[scale=0.18]{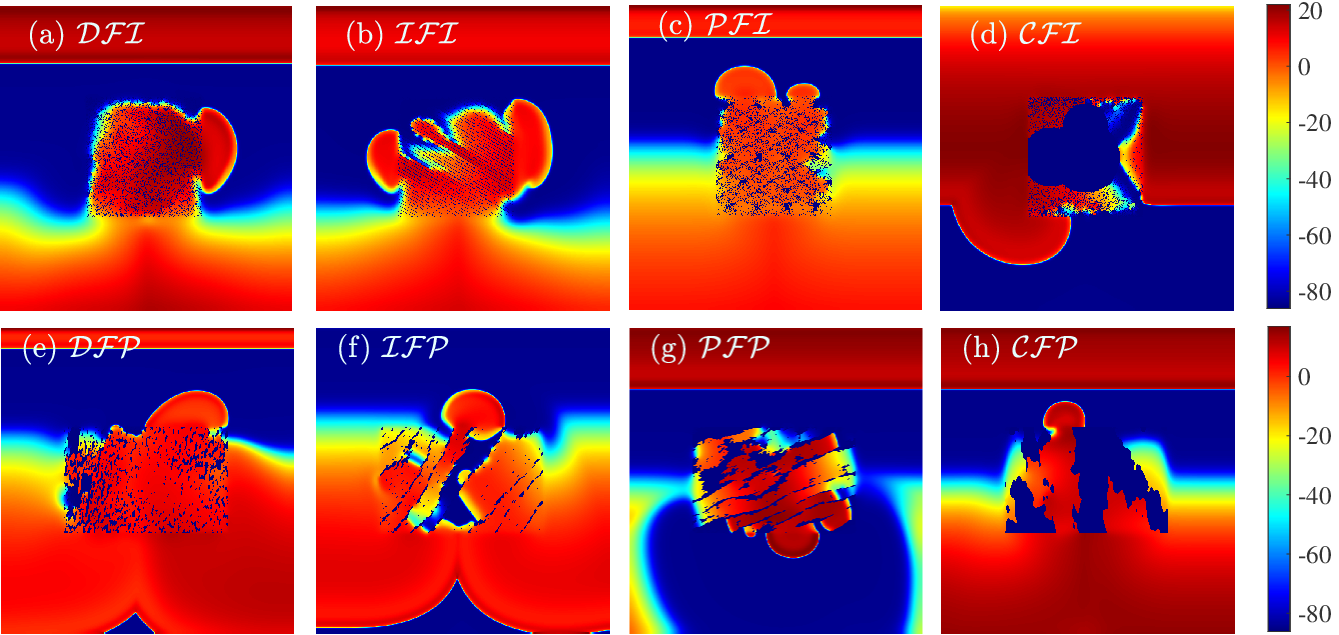}
\caption{(Color online) Illustrative pseudocolor plots of $V_m$ when we pace our simulation domain at its
lower boundary with a high-frequency ($\omega= 3.3$ Hz) current pulse.
(For the full spatiotemporal evolution of $V_m$ see the video V1 in the
Supplemental Material~\cite{Supp_Mat}.) Top row for the idealised model (B): (a) $\mathcal{DFI}$; (b) $\mathcal{IFI}$; (c) $\mathcal{PFI}$;
and (d) $\mathcal{CFI}$.  Bottom row for the Perlin-noise
model (A): (e)
$\mathcal{DFP}$; (f) $\mathcal{IFP}$; (g) $\mathcal{PFP}$; and (h)
$\mathcal{CFP}$.}
\label{fig:fig3}		
\end{figure}

\textcolor{black}{Arrhythmogenicity arises because of the interaction between the wave of electrical activation and the fibrotic tissue. The ratio of the wavelength of this wave and the linear size of the fibrotic tissue is an important control parameter Ref.~\cite{majumder2014turbulent, zimik2017reentry}}. 
We use the following simulation domains: (a) For the Perlin-noise model (A): a
square domain with $720 \times 720$ grid points; most of the
sites in these domains contain normal myocytes, except in a
central fibrotic region with $400 \times 250$ grid points. (b)
For the idealised model (B): a square domain with $512 \times
512$ grid points, with normal myocytes, except in a central
fibrotic region with $200 \times 200$  grid points. The area
fractions  of fibrotic regions are $\simeq 0.19$ and $\simeq
0.15$ in domains (a) and (b), respectively.
In our numerical simulations, we use fixed time and space steps $\Delta t = 0.02
\ ms$ and $\Delta x = 0.025 \ cm$, respectively, and a
finite-difference scheme, with a five-point stencil for the
Laplacian in Eq.~\ref{eq:VmPDE}. 
The value of $D=0.00154  \ cm^2/ms$ that
we use leads to the experimentally observed conduction velocity
$CV \simeq 70 \ cm/s$ in a region with normal myocytes~\cite{ten2006alternans}.	
\newline
A wave of electrical activation slows down in a region with RMs; this can lead
to conduction blocks that are arrhythmogenic. We pace our simulation domain at
its lower boundary with a high-frequency ($\omega= 3.3$ Hz) current pulse. The
resulting spatiotemporal evolution of $V_m$, given in the video V1 of the
Supplemental Material~\cite{Supp_Mat}, shows the birth of proto spirals, a
clear signature of arrhythmogenesis; we give illustrative pseudocolor plots of
$V_m$ in Fig.~\ref{fig:fig3} for $\mathcal{DF}$, $\mathcal{IF}$),
$\mathcal{PF}$, and $\mathcal{CF}$ in models (A) [top row] and (B) [bottom
row].

To quantify the statistical properties of these fibrotic-tissue patterns, we
first calculate their fractal dimensions $\mathbb{D}$ and lacunarity
$\mathcal{L}(\epsilon)$ (see, e.g.,
Refs.~\cite{delacalleja2020fractal,gould2011multifractal}), at a length scale
$\epsilon$, by using, respectively, the \textit{box-counting} and
\textit{gliding-box-counting}
algorithms~\cite{tolle2008efficient,gould2011multifractal}.
$\mathcal{L}(\epsilon)$, which measures the distribution of the
sizes of \textit{lacunae}, the degree of inhomogeneity, and translational and
rotational invariance of a pattern~\cite{gould2011multifractal,karperien2015fractal}, 
is given by
\begin{equation}
\mathcal{L}(\epsilon) = \frac{N(\epsilon) Q_{1}}{Q_{2}},
\label{eqn:lacunarity}
\end{equation}
where $N(\epsilon)$ is the number of square boxes of side $\epsilon$,
$p(i,\epsilon)$ the number of signal pixels in the $i^{th}$ box with
$i\in[1, N(\epsilon)]$, $Q_{1} \equiv \sum_{i} p(i,\epsilon)$, and $Q_{2} \equiv \sum_{i}
p(i,\epsilon)^2$. For the values of $\epsilon$ we use, we find, as in 
Ref.~\cite{gould2011multifractal}, that our data can be fit to the form
\begin{equation}
\mathcal{L}(\epsilon) = b \ \epsilon^{-a},
\label{eqn:Lac_hypr_fit}
\end{equation}  
where $b$ is the \textit{lacunarity parameter} and $a$ is the
\textit{lacunarity exponent} (see Figs. S4 and S5 in the Supplemental
Material~\cite{Supp_Mat}); a small value of $b$ leads to wide concavity in the
hyperbolic fit (see Eq.~\ref{eqn:Lac_hypr_fit}). We also characterize these 2D
fibrotic textures by their Betti numbers~\cite{delacalleja2020fractal}
$\beta_0$ and $\beta_1$, which measure, respectively, the number of connected
components and the number of holes that are completely enclosed by occupied
pixels (see Fig.S1 in the Supplemental Material~\cite{Supp_Mat}). To obtain
$\beta_0$ and $\beta_1$, we convert the fibrotic-tissue data sets into the
bit-map (bmp) image format; and then we use the computational-homology-project
software~\cite{CHomP} to get $\beta_0$ and $\beta_1$ for the particular
fibrotic image.  

\begin{figure}
\center
\includegraphics[scale=0.24]{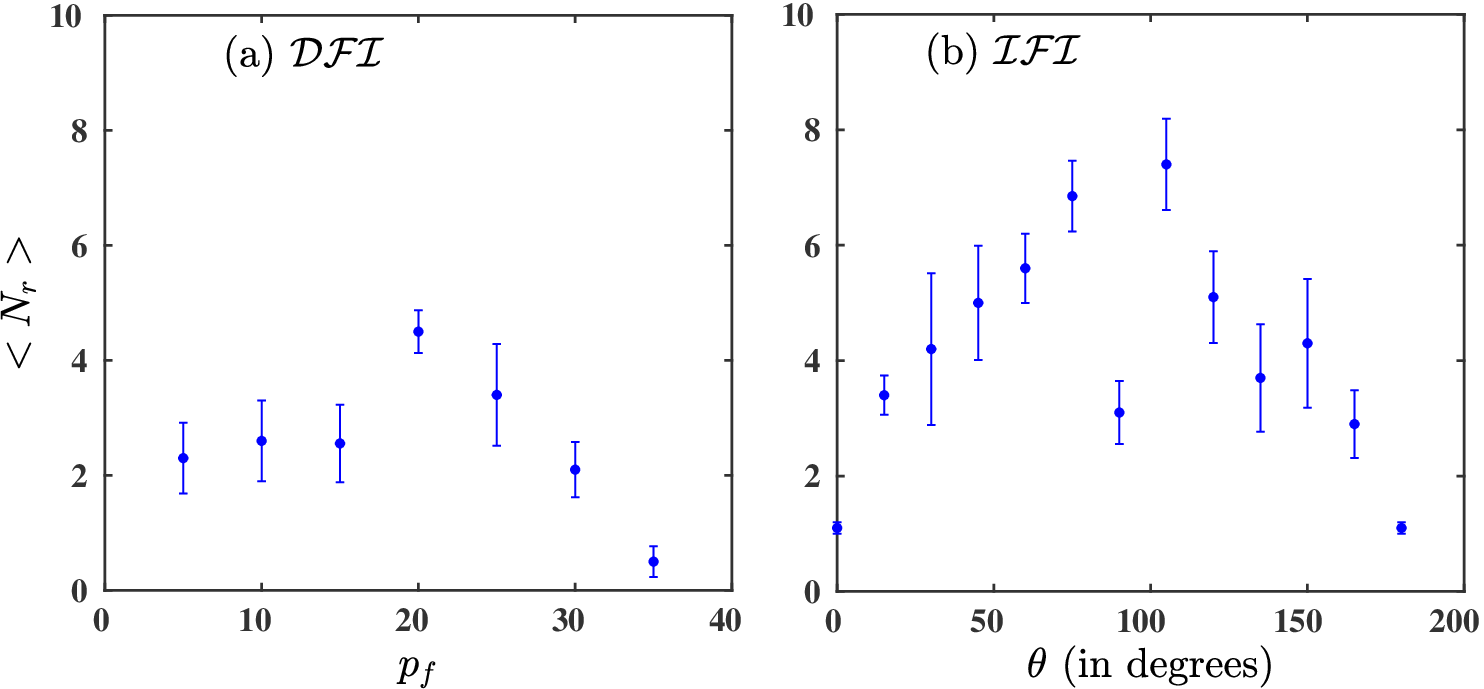}
\caption{(Color online) Plots of the mean number $\langle N_r \rangle$  of re-entries
that we observe, while pacing the fibrotic tissue with PCL=$300$ ms in the 
idealised model (B): (a) versus the percentage $p_f$ of randomly placed fibrotic
sites (in $\mathcal{DFI}$); (b) versus the angle $\theta$ of the inclination of 
the fibrotic strands with the pacing plane wave (in $\mathcal{IFI}$).}
\label{fig:fig4}
\end{figure}

In Fig.~\ref{fig:fig4}, we present plots of the mean number $\langle N_r
\rangle$  of re-entries that we observe, while pacing the fibrotic tissue with
PCL=$300$ ms in the idealised model (B); these data are averaged over $10$
realisations. The larger the value of $\langle N_r
\rangle$, the more arrhythmogenic this tissue. In Fig.~\ref{fig:fig4} (a) we
plot $\langle N_r \rangle$ versus $p_f$ for $\mathcal{DFI}$; for $p_f > 35 \%$,
there is no re-entry because of the very low conduction velocity of the
excitations within the fibrotic region; but $\langle N_r \rangle > 0$, for $0\%
< p_f \leq 30\%$, so $\mathcal{DFI}$ is clearly arrhythmogenic. In
Fig.~\ref{fig:fig4} (b) we show how $\langle N_r \rangle$, for $\mathcal{IFI}$,
depends on the angle $\theta$ of the inclination of the fibrotic strands with
the pacing plane wave; $\langle N_r \rangle$ is highest in the range 
$\theta \simeq 75^\circ - 105^\circ$ and it decreases outside this range. Clearly, 
$\theta$ is an important parameter which determines the arrhytmogenicity of
$\mathcal{IFI}$. We find that $\langle N_r \rangle$ in $\mathcal{PFI}$ is comparable
to that in $\mathcal{IFI}$. By contrast, $\mathcal{CFI}$ shows much lower values of
$\langle N_r \rangle$ than the other types of fibrotic patterns.

\begin{figure*}
 \center
\includegraphics[scale=0.248]{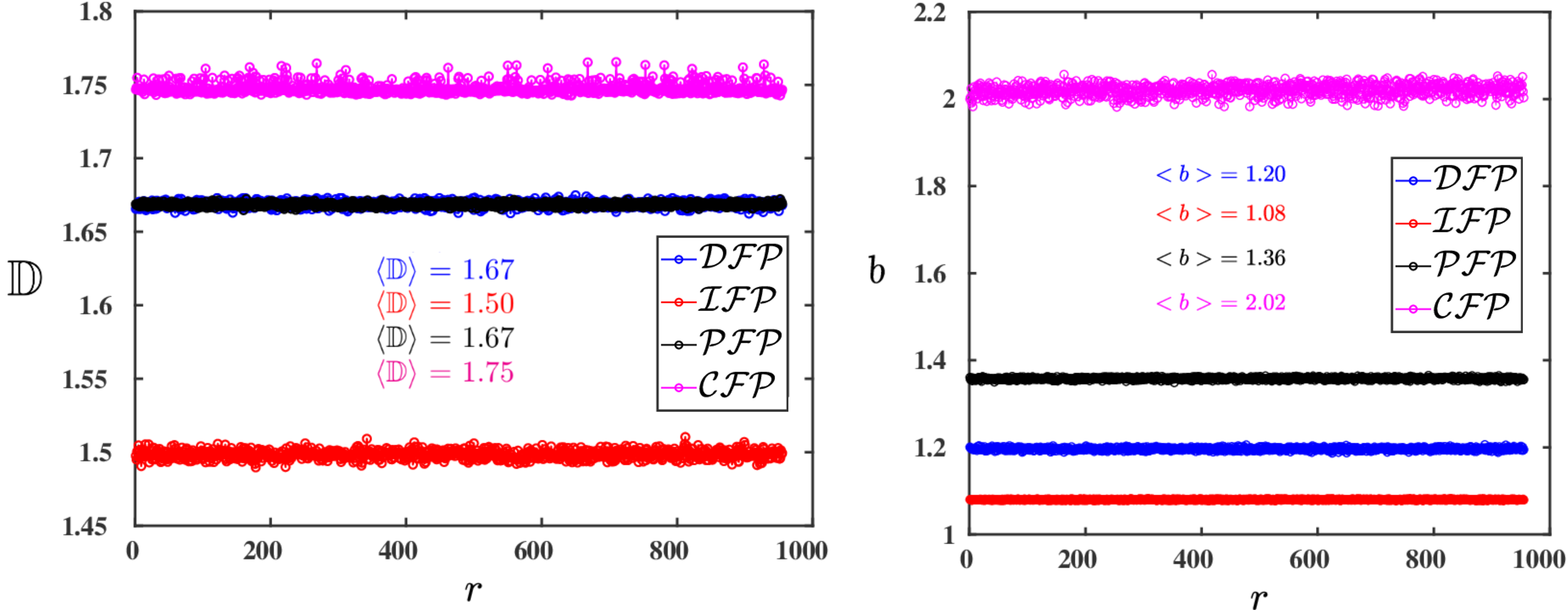}
\caption{(Color online) Plots versus the realisation number $r = 1, 2, \ldots, 955$, for the Perlin-noise
model (A), of the fractal dimension $\mathbb{D}$ (left panel) and the lacunarity
parameter $b$ (right panel) in Eq.~\ref{eqn:Lac_hypr_fit} fibrotic regions of types 
$\mathcal{DFP}$ (blue), $\mathcal{IFP}$ (red), $\mathcal{PFP}$ (black), and $\mathcal{CFP}$
(pink); we use angular brackets for mean values.}	
\label{fig:frac_lac}	
\end{figure*}

In Fig.~\ref{fig:frac_lac} we present, for the Perlin-noise model (A), 
plots versus the realisation number $r = 1, 2, \ldots, 955$ of the fractal dimension 
$\mathbb{D}$ (left panel) and the lacunarity parameter (right panel) $b$,
defined in Eq.~\ref{eqn:Lac_hypr_fit}, for all types of fibrotic regions, namely, 
$\mathcal{DFP}$ (blue), $\mathcal{IFP}$ (red), $\mathcal{PFP}$ (black), and $\mathcal{CFP}$
(pink); we use angular brackets for mean values. We see from these plots that
$\mathbb{D}$ is the same (within error bars) for $\mathcal{DFP}$ and $\mathcal{PFP}$;
however, these four different fibrotic patterns are distinguished clearly
by their lacunarity parameters $b$. 

\begin{figure*}
 \center
\includegraphics[scale=0.31]{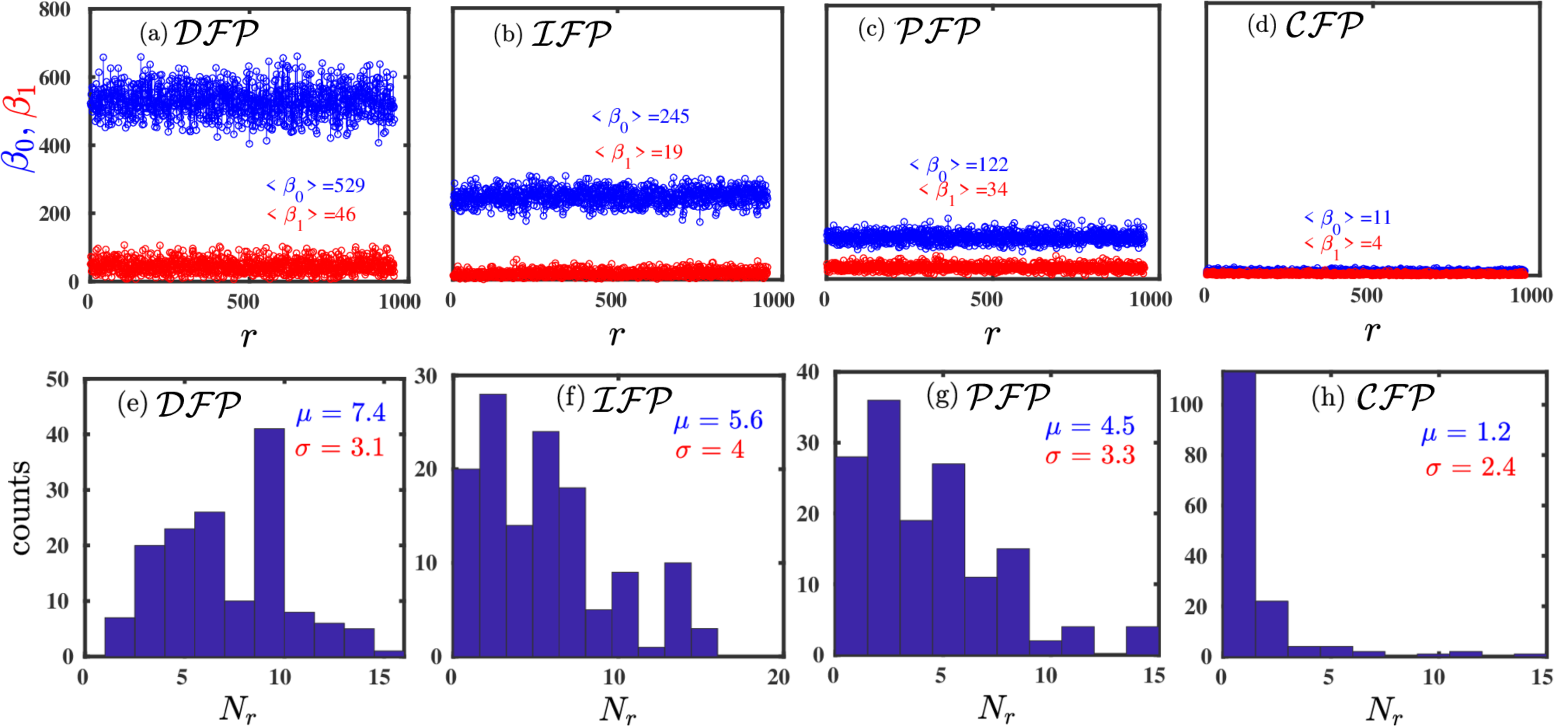}
\caption{(Color online) Plots versus the realisation number $r = 1, 2, \ldots, 955$, for the Perlin-noise
model (A), of the Betti numbers $\beta_0$ (blue) and $\beta_1$ (red) for
(a) $\mathcal{DFP}$, (b) $\mathcal{IFP}$, (c) $\mathcal{PFP}$, and (d) $\mathcal{CFP}$; we 
use angular brackets for mean values. Histograms of $N_r$, for $150$ model-(B) 
realizations of fibrotic regions, for (e) $\mathcal{DFP}$, (f) $\mathcal{IFP}$,
(g) $\mathcal{PFP}$, and (h) $\mathcal{CFP}$; $\mu$ and $\sigma$ are, respectively, 
the mean and standard deviation.}
\label{fig:Betti_numbers}			
\end{figure*}

In Fig.~\ref{fig:Betti_numbers}	we present, in the top row, plots versus the 
realisation number $r = 1, 2, \ldots, 955$, for the Perlin-noise
model (A), of the Betti numbers $\beta_0$ (blue) and $\beta_1$ (red) for
(a) $\mathcal{DFP}$, (b) $\mathcal{IFP}$, (c) $\mathcal{PFP}$, and (d) $\mathcal{CFP}$; we 
use angular brackets for mean values. In the bottom row of Fig.~\ref{fig:Betti_numbers}
we give histograms of $N_r$, which we obtain from $150$ model-(A) 
realizations of the fibrotic regions $\mathcal{DFP}$, $\mathcal{IFP}$,
$\mathcal{PFP}$, and $\mathcal{CFP}$; here, $\mu$ and $\sigma$ denote, respectively, 
the mean and standard deviation. 
	
\begin{figure*}
 \center
	\includegraphics[scale=0.33]{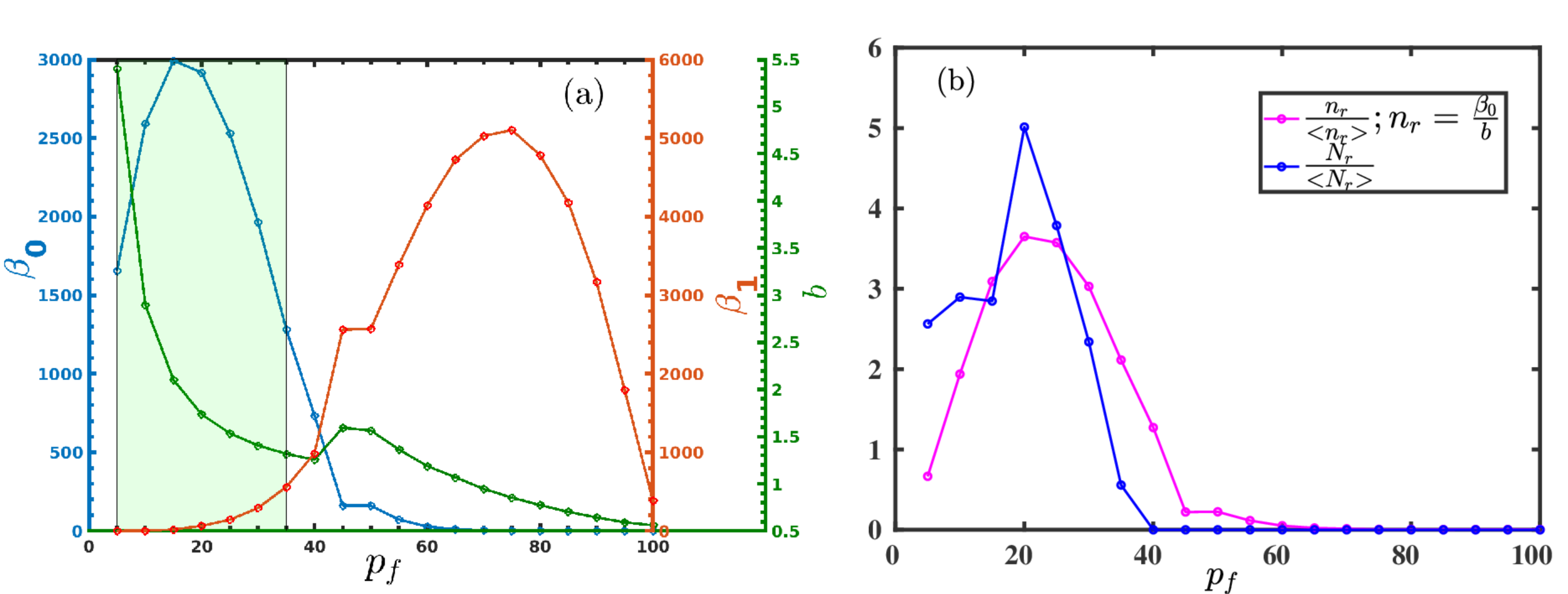}
\caption{(Color online) Plots versus $p_f$: (a) the Betti numbers $\beta_0$ (blue) and
$\beta_1$ (maroon) and the lacunarity parameter $b$ (green); the
light-green rectangle indicates the region in which $\langle N_r
\rangle$ is significant; (b) $N_r/\langle N_r \rangle$ (blue curve) and
$n_r/\langle n_r \rangle$ (pink curve).}	
\label{fig:risk_factor1}	
\end{figure*}

We conclude from Figs.~\ref{fig:frac_lac} and \ref{fig:Betti_numbers} that 
fibrotic patterns with small values of $b$ and with large values of $\beta_0$ 
are most arrhythmogenic. We show this explicitly for model-(B) 
realizations of $\mathcal{DFI}$ in Fig.~\ref{fig:risk_factor1}:
In Fig.~\ref{fig:risk_factor1}(a), we plot versus $p_f$ the Betti numbers 
$\beta_0$ (blue) and $\beta_1$ (maroon), and the lacunarity parameter $b$ (green);
the light-green rectangle indicates the region in which there is 
significant re-entry with a significant value of $\langle N_r \rangle$.
In Fig.~\ref{fig:risk_factor1}(b) we plot, versus $p_f$, $N_r/\langle N_r \rangle$ 
(blue curve) and $n_r/\langle n_r \rangle$ (pink curve), where $n_r \equiv \beta_0/b$;
these curves (blue and pink) are correlated to the extent that they are large in the 
same range of values of $p_f$.

Earlier computational studies of cardiac-tissue fibrosis include
Refs.~\cite{zlochiver2008electrotonic,xie2009effects,mcdowell2011susceptibility,nayak2013spiral,nayak2015turbulent,ten2007influence,majumder2012nonequilibrium,alonso2016reentry};
these are, roughly speaking, of two different types: (A) those that model the
myocyte-fibroblast
coupling~\cite{zlochiver2008electrotonic,xie2009effects,mcdowell2011susceptibility,nayak2013spiral,nayak2015turbulent};
and (b) those that use geometrical modeling for $\mathcal{DF}$
tissue~\cite{ten2007influence,majumder2012nonequilibrium,alonso2016reentry}.
There have been no studies, heretofore, which have investigated
arrhythmogenesis, systematically and simultaneously, in all four types of
fibrotic tissue. Our study leads to a natural way of quantifying the
arrhythmogenicity of diffuse fibrosis ($\mathcal{DF}$), interstitial fibrosis
($\mathcal{IF}$), patchy fibrosis ($\mathcal{PF}$), and compact fibrosis
($\mathcal{CF}$) in cardiac tissue. We have shown that the statistical
properties of these fibrotic-tissue patterns, such as their fractal dimension
$\mathbb{D}$, lacunarity parameter $b$, and Betti numbers $\beta_0$ and
$\beta_1$ are important in determining the arrhythmogenicity of fibrotic
tissue. Our work sets the stage for (a) to experimental investigations of
arrhythmogenesis in $\mathcal{DF}$, $\mathcal{IF}$, $\mathcal{PF}$, and
$\mathcal{CF}$ and (b) \textit{in silico} studies of such studies that go
beyond our model by using anatomically realistic simulation domains, with
muscle-fiber orientation, realistic myocyte-fibroblast couplings, and bidomain
models.

\begin{acknowledgments}

MKM and RP thank Jaya Kumar Alageshan for discussions, the Department of Science and 
Technology (DST), India, and the Council for Scientific and Industrial Research (CSIR), 
India, for financial support, and the Supercomputer Education and Research Centre 
(SERC, IISc) for computational resources. BAJL thank the Australian Research Council for 
financial support (Grant no:CE140100049).

\end{acknowledgments}

\end{document}


\title{Supplementary Material: Arrhythmogenicity of cardiac fibrosis: fractal measures  
and Betti numbers}
\author{Mahesh Kumar Mulimani}
\email{maheshk@iisc.ac.in ;}
\affiliation{Centre for Condensed Matter Theory, Department of Physics, Indian Institute of Science, Bangalore 560012, India.}
\author{Brodie A. J. Lawson}
\email{brodie.lawson86@gmail.com ;}
\affiliation{Centre for Data Science, Queensland University of Technology, Brisbane, Australia.}
\affiliation{ARC Centre of Excellence for Mathematical and Statistical Frontiers, Queensland University of Technology, Brisbane, Australia.}
\author{Rahul Pandit}
\email{rahul@iisc.ac.in}
\altaffiliation[\\]{also at Jawaharlal Nehru Centre For
Advanced Scientific Research, Jakkur, Bangalore, India}
\affiliation{Centre for Condensed Matter Theory, Department of Physics, Indian Institute of Science, Bangalore 560012, India.}

\maketitle

\begin{figure*}
	\centering
        \includegraphics[scale=0.45]{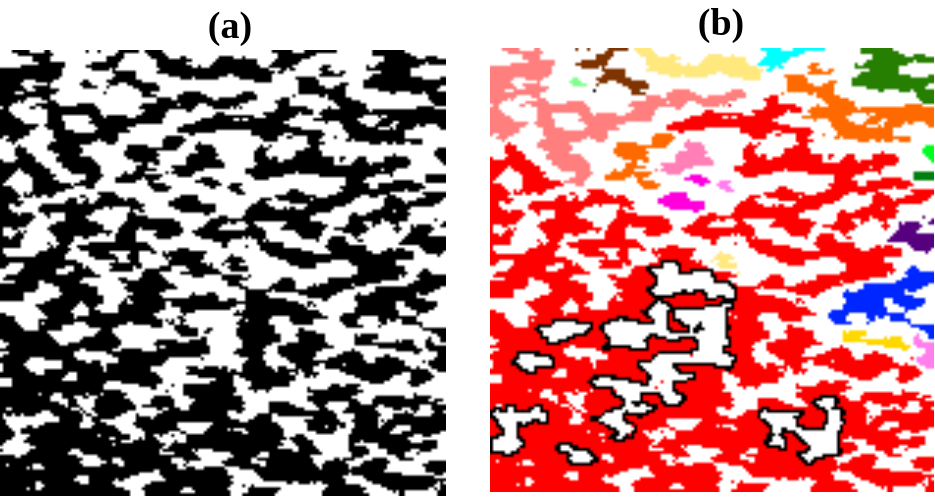}
\caption{(Color online) The computation of $\beta_0$ and $\beta_1$, which measure,
respectively, the number of connected components and the number of
holes that are completely enclosed by occupied pixels. To obtain
$\beta_0$ and $\beta_1$, we convert the fibrotic-tissue data sets
into the bit-map (bmp) image format; and then we use the
computational-homology-project software~\cite{CHomP} to get
$\beta_0$ and $\beta_1$ for the particular fibrotic image. (a) A
representative image for a region with diffuse fibrosis
$\mathcal{DFP}$ in the Perlin-noise model (A) [see the main paper].
(b) A colored version of the image in (a), with different colors for
the major, distinct, connected components; we have drawn black
boundaries for a few, representative,  holes, which are completely enclosed by
occupied pixels; by identifying such connected components and fully
enclosed holes, we obtain the Betti numbers; a careful computation
for this image yields $\beta_0=35$ and $\beta_1=72$.}				

	   \label{fig:figs1}
    \end{figure*}

\begin{figure*}[!htb]
	\centering
       \includegraphics[scale=0.40]{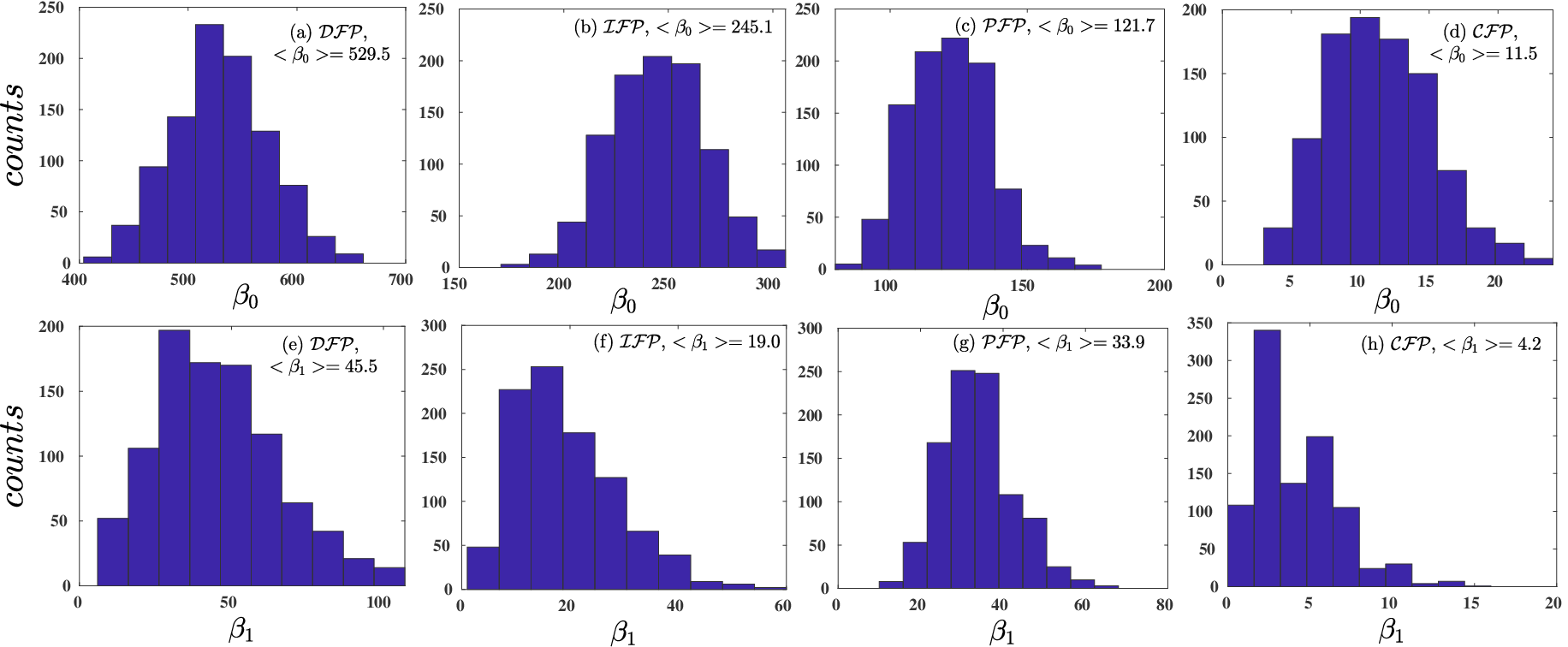}
	\caption{(Color online) Histograms of Betti numbers, from $955$ different 
fibrotic-tissue realizations for every fibrotic type.}	
\label{fig:figs2}
\end{figure*}

\begin{figure*}
       \includegraphics[scale=0.20]{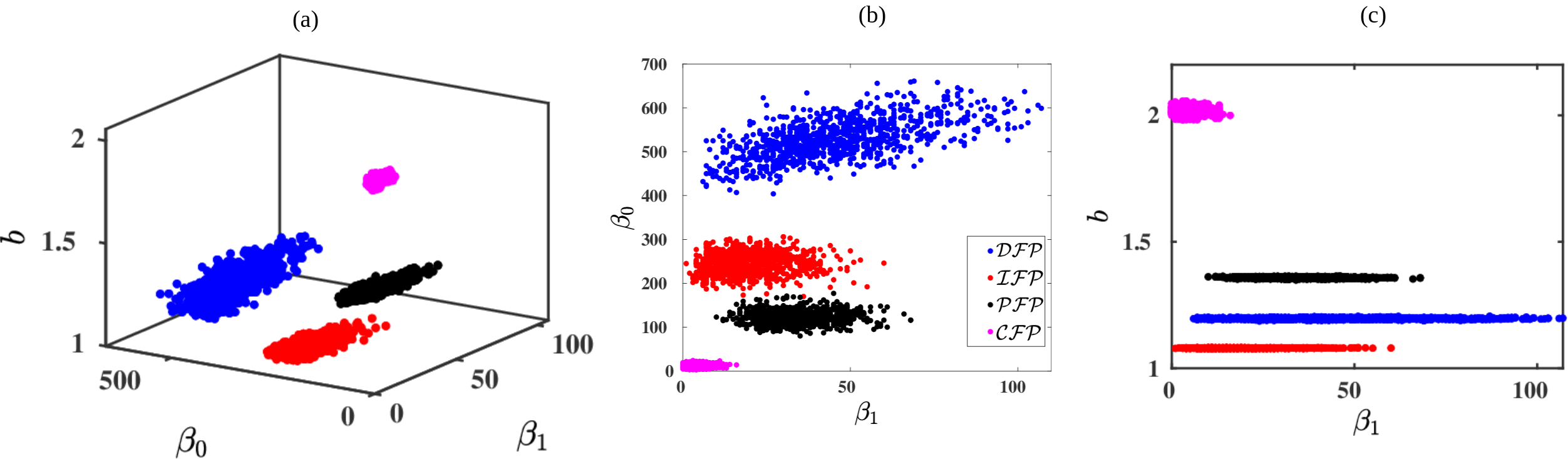}
\caption{(Color online) Scatter plots, in fibrotic tissue with diffuse fibrosis ($\mathcal{DFP}$),
interstitial fibrosis ($\mathcal{IFP}$), patchy fibrosis
($\mathcal{PFP}$), and compact fibrosis ($\mathcal{CFP}$); we use
the Perlin-noise~\cite{jakes2019perlin} model (A) (see the main
paper) and $955$ different realizations for each type of region, namely, 
$\mathcal{DFP}$, $\mathcal{IFP}$), $\mathcal{PFP}$, and $\mathcal{CFP}$: 
(a) the lacunarity parameter $b$ and the Betti numbers $\beta_0$, and $\beta_1$; 
(b) $\beta_0$ and $\beta_1$; and (c) $b$ and $\beta_1$.}
\label{fig:figs3}
\end{figure*}

\begin{figure*}
	\centering
        \includegraphics[scale=0.35]{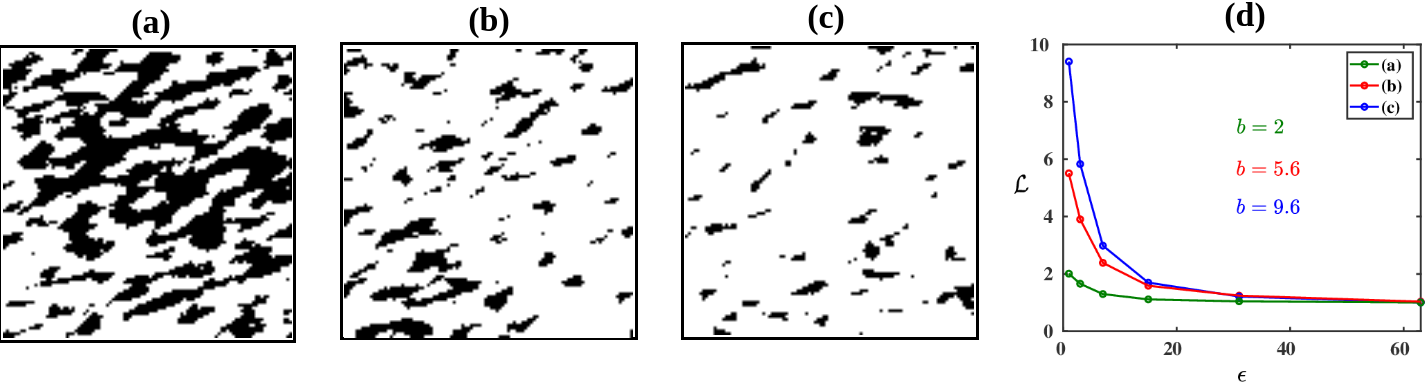}
	\caption{(Color online) Three representative images (a)-(c) of
diffuse fibrosis ($\mathcal{DFP}$); we use the Perlin-noise~\cite{jakes2019perlin} 
model (A) [see the main paper]. (d) Plots, for each one of the images (a)-(c), 
of the lacunarity $\mathcal{L}(\epsilon)$ versus the box length $\epsilon$ 
[see Eq. (3) in the main paper]; by using Eq. (4) in the main paper, we obtain
the lacunarity parameter $b$.}				
\label{fig:figs4}
    \end{figure*}

\begin{figure*}
	\centering
        \includegraphics[scale=0.30]{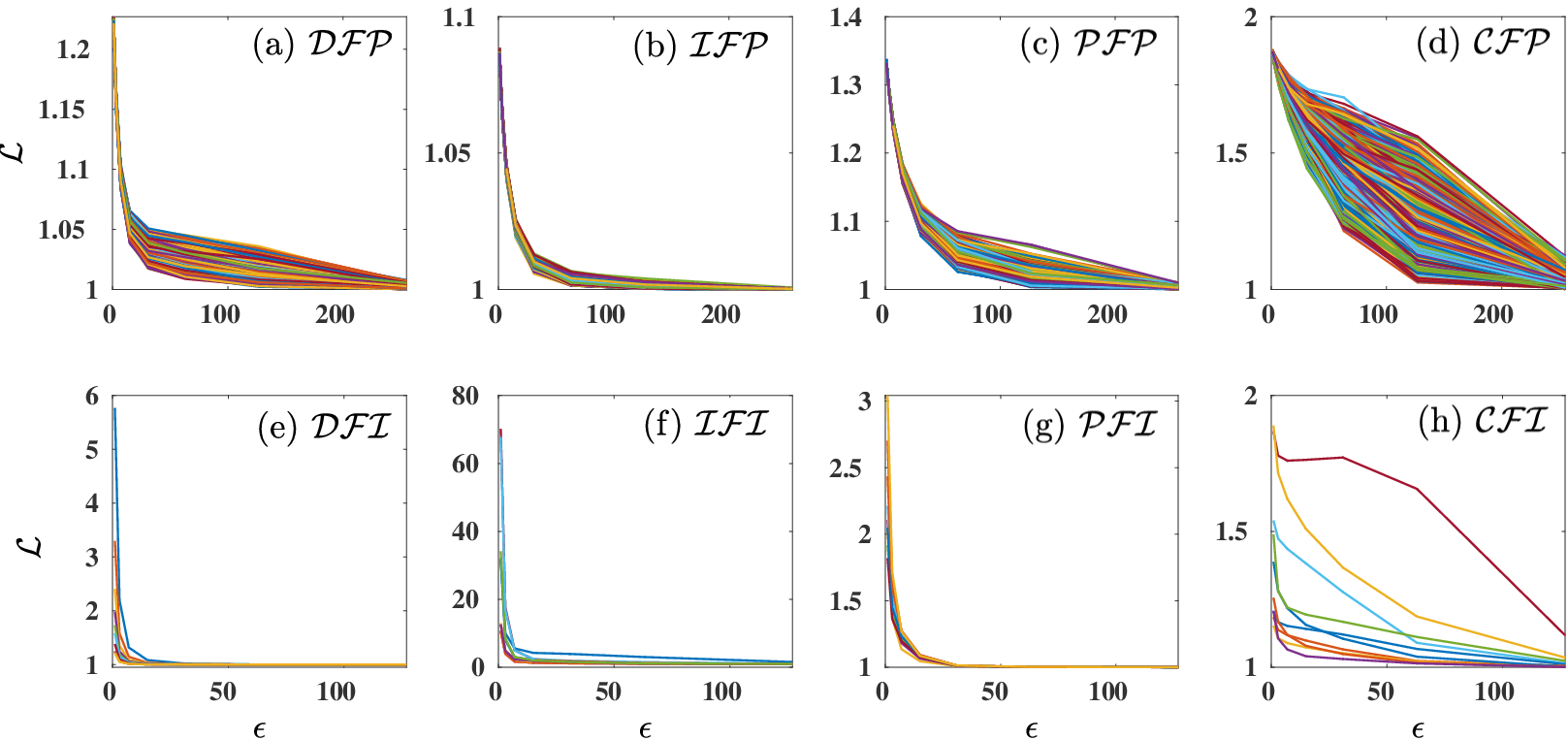}
\caption{(Color online) We assess the quality of the fit, given Eq.(4) in 
the main paper, by plotting 
the lacunarity $\mathcal{L}(\epsilon)$ versus the box length $\epsilon$ 
[see Eq. (3) in the main paper]. First row: for the Perlin-noise~\cite{jakes2019perlin} 
model (A) [see the main paper] for (a) $\mathcal{DFP}$, (b) $\mathcal{IFP}$, 
(c) $\mathcal{PFP}$, and (d) $\mathcal{CFP}$; second row: for the idealised 
model (B) [see the main paper] for (e) $\mathcal{DFI}$, (f) $\mathcal{IFI}$, 
(g) $\mathcal{PFI}$, and (h) $\mathcal{CFI}$. From such plots we obtain
the lacunarity parameter $b$; clearly, this fit is not a good one in some cases,
especially for $\mathcal{CFP}, \ \mathcal{CFI}$ (but, even for this case, the mean value
of $R^2$, the goodness of fit, is $81.4\%$).}
\label{fig:figs5}
\end{figure*}

\begin{figure*}
 \center
	\includegraphics[scale=0.25]{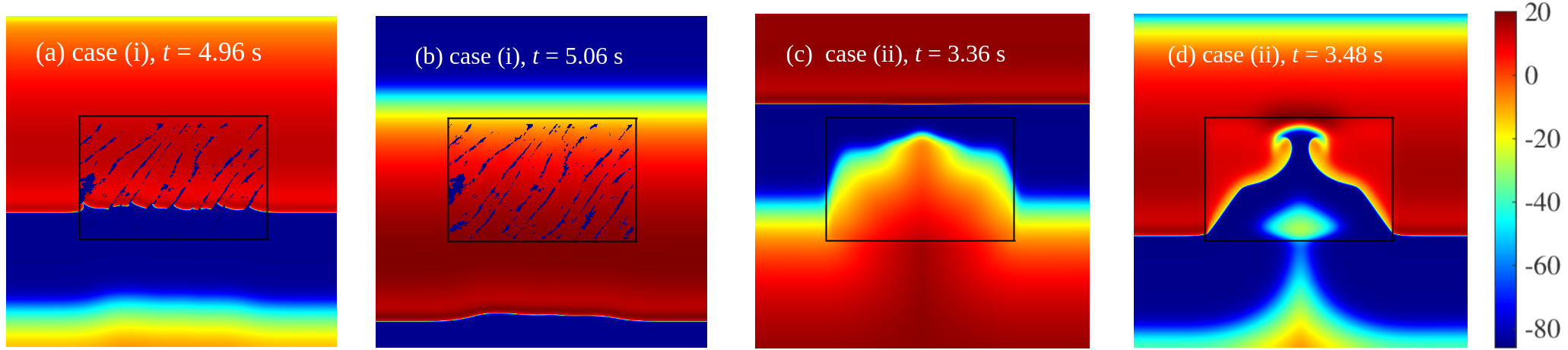}
	\caption{(Color online) Representative pseudocolor plots of the 
transmembrane potential $V_m$ which show that there is no arrhythmogenesis if
we use only (i) fibrotic tissue with no remodelling of myocytes or 
(ii) a remodelled region with no fibrotic tissue.}
\label{fig:figs6}	
\end{figure*}

We give below the following supplementary information, which is mentioned in the 
main paper:

\begin{enumerate}

\item We can characterize 2D fibrotic textures by their Betti 
numbers~\cite{delacalleja2020fractal} $\beta_0$ and $\beta_1$, which measure, 
respectively, the number of connected components and the number of holes that are 
completely enclosed by occupied pixels (see Fig.~\ref{fig:figs1}). To obtain
$\beta_0$ and $\beta_1$, we convert the fibrotic-tissue data sets into the
bit-map (bmp) image format; and then we use the computational-homology-project
software~\cite{CHomP} to get $\beta_0$ and $\beta_1$ for the particular
fibrotic image.  

\item In Fig.~\ref{fig:figs2} we present histograms of the Betti numbers
$\beta_0$ and $\beta_1$, in fibrotic tissue with diffuse fibrosis
($\mathcal{DFP}$), interstitial fibrosis ($\mathcal{IFP}$),
patchy fibrosis ($\mathcal{PFP}$), and compact fibrosis
($\mathcal{CFP}$); we use the
Perlin-noise~\cite{jakes2019perlin} model (A) [see the main paper]
and $955$ different realizations for each type of region,
namely, $\mathcal{DF}$, $\mathcal{IF}$, $\mathcal{PF}$, and
$\mathcal{CF}$.

\item In Fig.~\ref{fig:figs3} we present the following scatter plots, in
fibrotic tissue with diffuse fibrosis ($\mathcal{DFP}$),
interstitial fibrosis ($\mathcal{IFP}$), patchy fibrosis
($\mathcal{PFP}$), and compact fibrosis ($\mathcal{CFP}$); we use
the Perlin-noise~\cite{jakes2019perlin} model (A) [see the main
paper] and $955$ different realizations for each type of region, namely, 
$\mathcal{DFP}$, $\mathcal{IFP}$, $\mathcal{PFP}$, and $\mathcal{CFP}$: 
(a) the lacunarity parameter $b$ and the Betti numbers $\beta_0$, and $\beta_1$; 
(b) $\beta_0$ and $\beta_1$; and (c) $b$ and $\beta_1$.

\item In Figs.~\ref{fig:figs4} (a)-(c) we present three representative images  of
diffuse fibrosis ($\mathcal{DFP}$); we use the Perlin-noise~\cite{jakes2019perlin} 
model (A) [see the main paper]. In Fig.~\ref{fig:figs4} (d) we plot, for each one of
the images, the lacunarity $\mathcal{L}(\epsilon)$ versus the box length $\epsilon$ 
[see Eq. (3) in the main paper], and, by using Eq. (4) in the main paper, we obtain
the lacunarity parameter $b$.

\item In Fig.~\ref{fig:figs5} we assess the quality of the fit, given Eq.(4) in 
the main paper, by plotting 
the lacunarity $\mathcal{L}(\epsilon)$ versus the box length $\epsilon$ 
[see Eq. (3) in the main paper]. First row: for the Perlin-noise~\cite{jakes2019perlin} 
model (A) [see the main paper] for (a) $\mathcal{DFP}$, (b) $\mathcal{IFP}$, 
(c) $\mathcal{PFP}$, and (d) $\mathcal{CFP}$; second row: for the idealised 
model (B) [see the main paper] for (e) $\mathcal{DFI}$, (f) $\mathcal{IFI}$, 
(g) $\mathcal{PFI}$, and (h) $\mathcal{CFI}$. From such plots we obtain
the lacunarity parameter $b$; clearly, this fit is not very good in some cases,
especially for $\mathcal{CFI}, \mathcal{CFP}$ (but, even for this case, the mean value
of $R^2$, the goodness of fit, is $81.4\%$).

\item In Fig.~\ref{fig:figs6} we present representative pseudocolor plots of the 
transmembrane potential $V_m$ which show that there is no arrhythmogenesis if
we use only (i) fibrotic tissue with no remodelling of myocytes or 
(ii) a remodelled region with no fibrotic tissue.

\end{enumerate}


%

